\documentclass{PoS}
\usepackage{wrapfig}
\usepackage{amsmath}
\title{Upward-Pointing Cosmic-Ray-like Events Observed with ANITA}

\ShortTitle{Upward-Pointing Cosmic-Ray-like Events Observed with ANITA}

\author{\speaker{Andres Romero-Wolf}%
        \\
       Jet Propulsion Laboratory, California Institute of Technology\\
       E-mail: \email{Andrew.Romero-Wolf@jpl.nasa.gov}}

\author{
(ANITA Collaboration) 
P. W. Gorham$^1$, 
J. Nam$^2$, 
S. Hoover$^4$,
P. Allison$^{5,6}$ 
O. Banerjee$^5$,
L. Batten$^{11}$,
J. J. Beatty$^{5,6}$,
K. Belov$^3$, 
D. Z. Besson$^7$, 
W. R. Binns$^8$, 
V. Bugaev$^8$, 
P. Cao$^9 $,
C. Chen$^2 $,
P. Chen$^2 $,
J. M. Clem$^9 $,
A. Connolly$^{5,6}$, 
B. Dailey$^5 $,
C. Deaconu$^{10}$,
L. Cremonesi$^{11} $,
P. F. Dowkontt$^4 $,
M. A. DuVernois$^1$, 
R. C. Field$^{12}$, 
B. D. Fox$^1 $,
D. Goldstein$^{14}$, 
J. Gordon$^5 $,
C. Hast$^{12}$,
C. L. Hebert$^1$, 
B. Hill$^1 $,
K. Hughes$^5 $,
R. Hupe$^5 $,
M. H. Israel$^8$, 
A. Javaid$^9 $,
J. Kowalski$^1$, 
J. Lam$^4 $,
A. Ludwig$^{10}$,
J. G. Learned$^1$,
K. M. Liewer$^3$, 
T. C. Liu$^2 $,
J. T. Link$^1 $,
E. Lusczek$^{15} $,
S. Matsuno$^1 $,
B. C. Mercurio$^5$, 
C. Miki$^1 $,
P.Mio\v{c}inovi\'c$^1$, 
M. Mottram$^{11}$,
K. Mulrey$^9 $,
C. J. Naudet$^3$,
 J. Ng$^{12} $,
R. J. Nichol$^{11}$,
A. Novikov$^7$, 
K. Palladino$^5 $,
S. Prohira$^7$, 
B. F. Rauch$^8$, 
K. Reil$^{13} $,
J. Roberts$^1 $,
M. Rosen$^1 $,
B. Rotter$^1$,
J. Russell$^1 $,
L. Ruckman$^1 $,
D. Saltzberg$^4$, 
D. Seckel$^9 $,
S. Stafford$^5$, 
J. Stockham$^7$,
M. Stockham$^7$,
B. Strutt$^{12}$, 
K. Tatem$^1$,
G. S. Varner$^1$,
A. G. Vieregg$^{10}$, 
D. Walz$^{12}$,
S. A. Wissel$^{15}$, 
and F. Wu$^4$
\\ 
$^1$University of Hawaii, Manoa, 
$^2$National Taiwan University, 
$^3$Jet Propulsion Laboratory, 
$^4$University of California, Los Angeles
$^5$Ohio State University, 
$^6$Center for Cosmology and Particle Astrophysics, Ohio State University, 
$^7$University of Kansas, Lawrence, 
$^8$Washington University in St. Louis, 
$^9$University of Delaware, Newark, 
$^{10}$Department of Physics, Enrico Fermi Institute, Kavli Institute for Cosmological Physics, University of Chicago, 
$^{11}$University College London, 
$^{12}$SLAC National Accelerator Laboratory, 
$^{13}$University of California, Irvine, 
$^{14}$University of Minnesota, Minneapolis, Minnesota, 
$^{15}$California Polytechnic State University, San Luis Obispo
}

\author{in collaboration with J. Alvarez-Muñiz$^{16}$, W. Carvalho Jr.$^{17}$, H. Schoorlemmer$^{18}$, E. Zas$^{16}$\\
        $^{16}$ Instituto Galego de Física de Alatas Enerx\`ias, University of Santiago de Compostela, $^{17}$Universidade de S\~ao Paulo, $^{18}$Max-Planck-Institut f\"ur Kernphysik}

\abstract{These proceedings address a recent publication by the ANITA collaboration of four upward-pointing cosmic-ray-like events observed in the first flight of ANITA. Three of these events were consistent with stratospheric cosmic-ray air showers where the axis of propagation does not intersect the surface of the Earth. The fourth event was consistent with a primary particle that emerges from the surface of the ice suggesting a possible $\tau$-lepton decay as the origin of this event. These proceedings follow-up on the modeling and testing of the hypothesis that this event was of $\tau$ neutrino origin.}

\FullConference{35th International Cosmic Ray Conference\\
                 10-20 July, 2017\\
                 Bexco, Busan, Korea}

\begin{document}


\section{Introduction}



The ANITA collaboration recently reported the observation of four upward-pointing cosmic-ray-like events observed during the first flight~\cite{Gorham_2016}. Three of these events were consistent with cosmic-ray events above the geometric horizon but below the horizontal. These stratospheric air showers belong to a new class of events that has not been previously observed. The fourth upward-pointing air shower was consistent with a primary particle exiting the surface of the Earth. Such events can arise from a $\tau$ lepton resulting from a $\tau$ neutrino ($\nu_\tau$) propagating through the Earth. The $\tau$ lepton exits the surface of the Earth and subsequently decays in the atmosphere producing an extensive air shower. 

The competing hypotheses for the $\tau$-lepton air shower candidate are that it is either a down-going cosmic ray extensive air shower with its radio impulsive emission reflected off the surface of the ice (see~\cite{Hoover_2010}) or an anthropogenic background (see~\cite{ANITA_instrument_paper, ANITA1_results, ANITA2_results}). The $\tau$-lepton decay is favored over the down-going cosmic ray hypothesis due to the polarity of the impulse and the correlation of the polarization vector to the geomagnetic field. The polarity of the impulse is flipped upon reflection due to the dielectric contrast of the atmosphere and the surface of the Antarctic ice cap and to the predominantly horizontal polarization of the impulse. The polarization vector is determined by the direction of the geomagnetic field at air shower maximum and the direction of the shower axis. The upward-going $\tau$ lepton and reflected down-going cosmic ray hypotheses each predict a different location of air shower maximum resulting in different predictions for the polarization. The radio impulse shape, polarization, and air shower geometry of these events is discussed in detail in~\cite{Gorham_2016}. 

Anthropogenic events that trigger the ANITA payload are unpredictable but readily identifiable by their tendency to cluster with each other. Although the $\tau$-lepton candidate event was isolated, there is always a chance an anthropogenic event may be isolated. The probability that an event of anthropogenic origin is consistent with an upward-going $\tau$-lepton event is based on the analysis of $\sim$80,000 anthropogenic events found during the first flight of ANITA. Using this population of anthropogenic events we estimated the fraction that have waveforms consistent with the population of extensive air shower events and the fraction of events where the polarization vector is consistent with the geomagnetic origin of the event. The relative probability of a $\tau$-lepton air shower to an anthropogenic origin is $\sim550$, which favors the former hypothesis but does not exclude the latter with high confidence. The analysis described here is discussed in more detail in ~\cite{Gorham_2016}. The data recently obtained with the third and fourth flights of ANITA, with ongoing blind analyses designed to address this disambiguation, will potentially confirm or falsify the $\tau$-lepton decay hypothesis. 

Although a $\nu_\tau$ origin of this event provides an attractive hypothesis for explaining the upward-pointing nature of the $\tau$-lepton decay candidate event, there are some difficulties with regard to other observational predictions. In the original publication~\cite{Gorham_2016} it was pointed out that the emergence angle of the event is $25.4^{\circ}$ with a $\sim 1^\circ$ uncertainty, meaning that the trajectory through the Earth has an attenuation coefficient of $4\times10^{-6}$ at 1~EeV, making it unlikely that such an event could have been observed with ANITA's exposure. In addition, standard model (SM) neutrino interaction cross-sections and $\tau$ lepton energy loss rates should lead to many more of these events observed closer to the horizon. The caveat is that even within the standard model, there are factors of 3-5 uncertainty in the neutrino cross-section at ultra-high energies in addition to other models beyond the SM that can further suppress or enhance the cross-section.

In these proceedings we present more details on the modeling of the acceptance of $\tau$-lepton air showers of $\nu_\tau$ origin. The process is complex due to the dependence of the air shower the radio emission on the index of refraction of air where the shower is developing. The full range of emergence angles and decay altitudes needed for an accurate estimate of the exposure requires an extensive set of air shower simulations that is currently in progress. For these proceedings we estimate an upper bound to the ANITA exposure, which allows us to make simplifying assumptions while providing a result that can be compared to other experiments. In these proceedings we present the ANITA $\nu_\tau$ acceptance model for the $\tau$-lepton decay channel, including a description of the $\tau$ lepton and neutrino propagation through the Earth, accounting for the effect of regeneration, the radio emission model applied, and the ANITA detector model used to produce upper bounds on the exposure. We compare these bounds with Auger and IceCube exposures. 
%
%

\section{Tau Neutrino Acceptance Model}
\begin{wrapfigure}[23]{r}{0.35\textwidth}
  \centering
   \includegraphics[width=\linewidth]{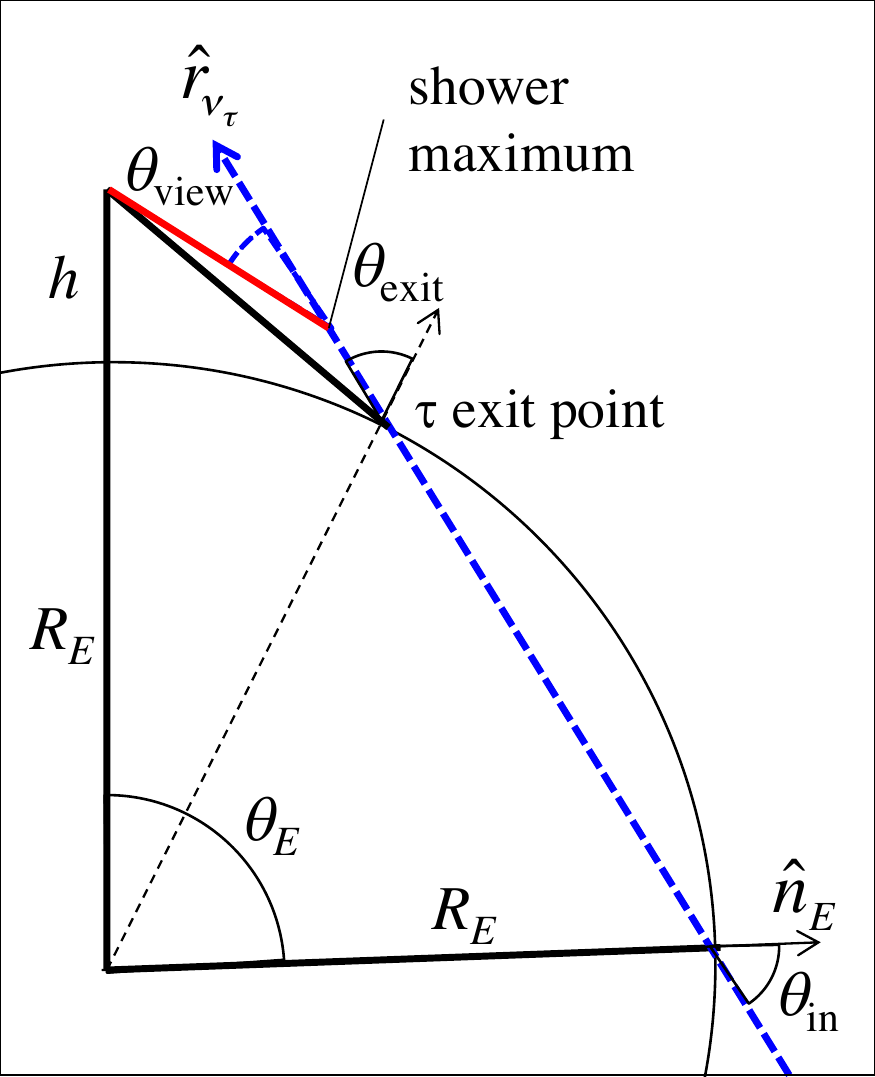} 
   \caption{Detection geometry. The blue dashed line represents the neutrino axis of propagation with direction $\hat{r}_{\nu_\tau}$ going through the Earth and piercing the surface of the Earth with the $\tau$ lepton exit point at latitude $\theta_E$. For a shower resulting from a $\tau$-lepton decay in the atmosphere, the maximum radio emission corresponds to the shower maximum, which is viewed by the detector at altitude $h$ with angle $\theta_{\mbox{\scriptsize view}}$.}
   \label{fig:geom}
\end{wrapfigure}

The $\nu_\tau$ acceptance takes the surface of the Earth's crust as the reference area. At each point on that surface, given by the angles $(\theta_E, \ \phi_E)$ in spherical coordinates, all directions of inward-bound neutrinos $\hat{r}_{\nu}$ are considered. The acceptance takes into account the projected area term $\hat{r}_{\nu}\cdot \hat{n}_E$, where $\hat{n}_E$ is the normal to the surface of the Earth at neutrino entry point. Given the $\nu_\tau$ incidence angle $\theta_\mathrm{in}$, we propagate the particle to estimate the probability that a $\tau$ lepton exits the surface of the Earth's crust at angle $\theta_\mathrm{exit}$. Note that $\theta_\mathrm{exit}=\theta_\mathrm{in}$ for a spherical Earth model. We consider all the outcomes determined by the probability $p_\mathrm{exit}(E_\tau|E_{\nu_\tau}, \theta_\mathrm{in})$ that a $\tau$ lepton exits with energy $E_\tau$ given the injected neutrino energy $E_{\nu_\tau}$ and incidence angle $\theta_\mathrm{in}$. After the $\tau$-lepton exits, the probability that it decays  
$p_\mathrm{decay}(t|E_\tau)$ in time $t$ given the energy $E_\tau$ is exponentially distributed. 
The $\tau$-lepton energy and location of the $\tau$-lepton decay $\vec{x}_\mathrm{decay}(t)$ determines the air shower and radio emission. The probability that ANITA detects the radio impulse is given by $p_\mathrm{det}(E_{\tau}, \vec{x}_\mathrm{decay}(t))$. The net sum of these effects results in the acceptance, obtained from the integral below:
\begin{equation}
\begin{split}
A_{\nu_\tau}(E_{\nu_\tau})= & R_E^2 \iint d\Omega_E \iint d\Omega_{\nu_\tau} \ \hat{r}_{\nu_\tau}\cdot \hat{n}_E 
\\ 
& 
 \ \ \ \int dE_{\tau } \ p_{\mbox{\scriptsize exit}}(E_{\tau}|E_{\nu_\tau}, \theta_{\mbox{\scriptsize exit}}) 
\\ 
&
\ \ \ \int dt  \ p_\mathrm{decay}(t|E_\tau) \ 
p_{\det}(E_{\tau}, \vec{x}_\mathrm{decay}(t)).
\end{split}
\end{equation}

The integral is evaluated using Monte Carlo sampling of the neutrino injection points, neutrino direction angles, the resulting $\tau$-lepton energy $E_\tau$ and decay times in the atmosphere. The models that go into the Monte Carlo estimate are described in the next section.


\section{Tau Neutrino Propagation}
We summarize here the $\nu_\tau$ and $\tau$ lepton propagation simulations used in this analysis (see~\cite{Alvarez-Muniz_2017} for more details). When the $\nu_\tau$ is injected, either a charged current (CC) or neutral current (NC) interaction will take place with the relative probability given by the ratio of the cross-sections ($\sigma_{CC}/\sigma_{NC}\sim3$). In case of a NC interaction, a hadronic jet along with a $\nu_\tau$ are produced. The $\nu_\tau$ energy is reduced by $\sim 0.8$ on average, and continues to propagate. In case of CC interaction a hadronic shower along with a $\tau$ lepton are produced. The $\tau$ lepton carries $\sim0.8$ of the parent $\nu_\tau$ energy, on average, and continues to propagate. 

As the $\tau$ lepton propagates it loses energy due to interactions with the surrounding matter and  eventually decays. If the decay occurs inside the Earth, it always results in a $\nu_\tau$ that continues to propagate with a fraction of the energy of the $\tau$ lepton. This process is known as ``regeneration". If the $\tau$ lepton exits the crust of the Earth, the energy losses in the atmosphere become negligible compared to the Earth due to the factor of $\sim 1000$ reduction in density. The $\tau$ lepton may decay in the atmosphere producing an extensive air shower detectable by ANITA or escape to space.

In this study, we have considered uncertainties of the neutrino interaction cross-section and $\tau$-lepton energy loss rates within the standard model. For the neutrino cross-section, we included the lower and upper ranges from~\cite{Connolly_2011} along with the middle (or standard) values, also from~\cite{Connolly_2011} as guiding examples of uncertainties within the SM. For the $\tau$-lepton energy loss rates, we have used the standard values of~\cite{ALLM_97} and a saturated model~\cite{ASW_2005}, which has generally lower loss values.


\section{Radio Emission Model}
\begin{wrapfigure}[22]{r}{0.6\textwidth}
  \centering
   \includegraphics[width=\linewidth]{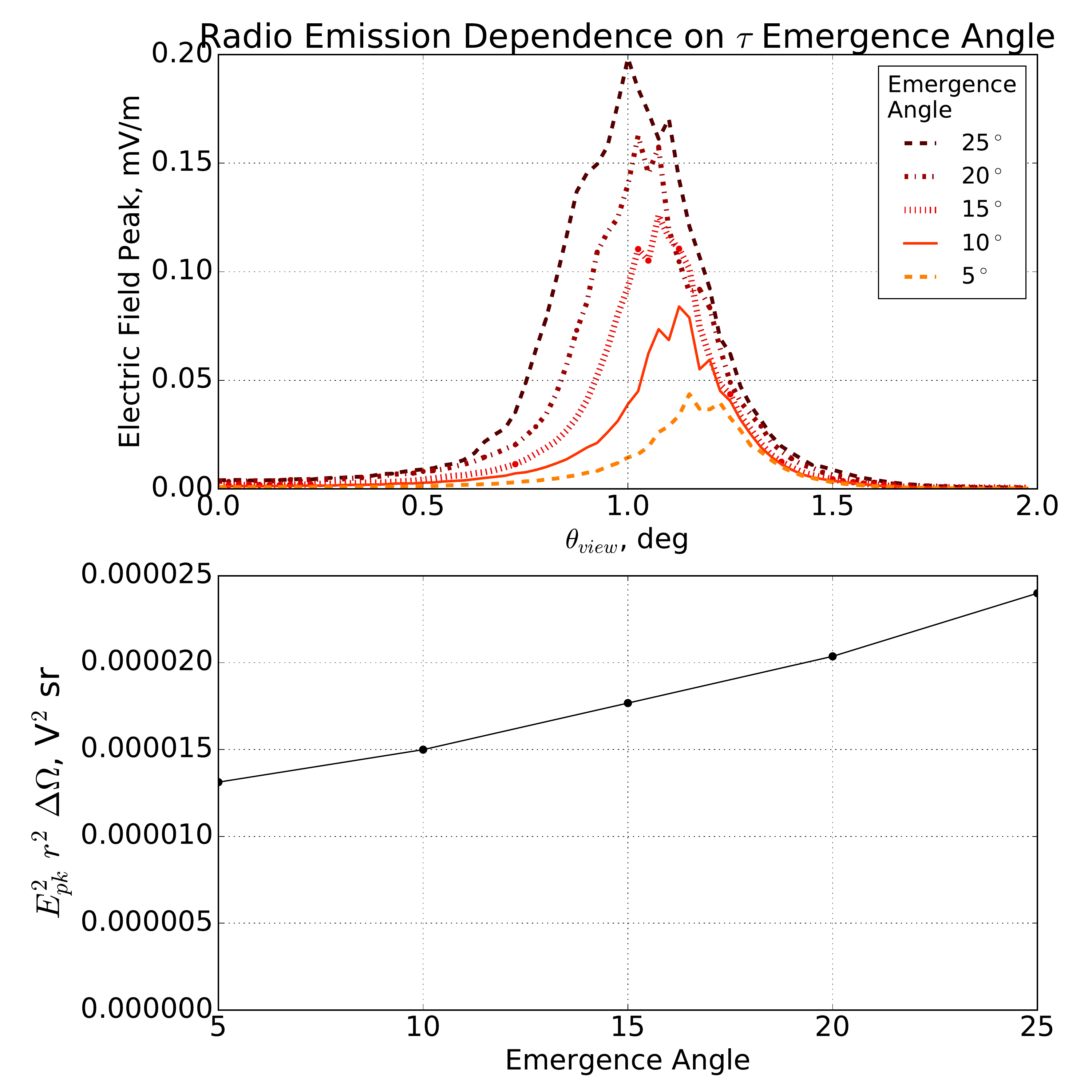} 
   \caption{The radio emission dependence on the $\tau$ emergence angle for a decay altitude of 0~km (see text for details).}
   \label{fig:tau_emergence}
\end{wrapfigure}

The radio emission of upward-pointing showers was simulated using ZHAireS~\cite{ZHAireS}. Given that the radio impulse depends on both the emergence angle and decay altitude of the $\tau$ lepton, resulting in the need for a large number of simulations, we have decided to find the relevant geometries with the highest radio emission power. In the top panel of Figure~\ref{fig:tau_emergence} we show the electric field peak in the 200-1200 MHz band for a $10^{17}$~eV hadronic shower as a function of the view angle $\theta_\mathrm{view}$ for various emergence angles with the $\tau$ lepton decay altitude set to $0$~km above the Antarctic ice sheet. We used the magnetic field at the location of the candidate event from~\cite{Gorham_2016}. In the bottom panel of Figure~\ref{fig:tau_emergence}, we estimate the square of the electric field peak times the distance to the ANITA payload at 37~km altitude multiplied by the solid angle corresponding to the full-width half max of $\theta_\mathrm{view}$. This quantity allows us to determine that the emergence angle of 25$^{\circ}$ has the highest amount of total radiated power of the simulated angles. Similarly, we determined that for various decay altitudes, at a fixed emergence angle of 25$^\circ$, the highest radiated power corresponds to decays at an altitude of 0~km above the Antarctic ice sheet.

We take the radio emission at a $\tau$ lepton decay altitude of 0~km above the Antarctic ice sheet with an emergence angle of 25$^\circ$ as our fiducial radio emission model. A parameterization that rescales the electric field peak $\mathcal{E}$ profile for the distance of payload to $\tau$-lepton decay $r$, view angle $\theta_\mathrm{view}$, $\tau$-lepton energy $E_\tau$ is given by

\begin{equation}
\epsilon(\theta_{view}) = \mathcal{E}_0\left[f\exp\left(-\frac{(\theta_{view} - \theta_{pk})^2}{2\sigma_{view}^2}\right) + (1-f)\left(1+\left(\frac{\theta_{view} - \theta_{pk}}{\sigma_{view}}\right)^2\right)^{-1}\right] + \mathcal{E}_1\exp\left(-\frac{\theta_{view}^2}{2\Sigma_{view}^2}\right) 
\end{equation}
and the electric field is
\begin{equation}
\mathcal{E}(E_{\tau}, r, \theta_{view}) = \left(\frac{E_{\tau}}{10^{17}\mbox{ eV}}\right)\left(\frac{r}{86.4\mbox{ km}}\right)\epsilon(\theta_{view})
\end{equation}
with best fit parameters $\mathcal{E}_0=0.183$~mV/m,  $\theta_{pk}=1.011^{\circ}$, $\sigma_{view}=0.16^{\circ}$, $f=0.825$, $\mathcal{E}_1=0.004$~mV/m, $\Sigma_{view}=1.135^{\circ}$.

\section{Detector Model}
The ANITA detector has a complex triggering system~\cite{ANITA_instrument_paper} requiring fairly sophisticated simulations to model it properly. Such simulations have been applied to neutrino~\cite{ANITA1_results, ANITA2_results} and cosmic ray~\cite{Hoover_2010, Schoorlemmer_2015} analyses. In this study, we take a simplified approach with the objective of setting an upper bound to the sensitivity of ANITA. 

Given the peak electric field $\mathcal{E}_\mathrm{pk}$, obtained from the simulations described in the previous section, we convert it to a peak voltage using 
\begin{equation}
\mathrm{V}_\mathrm{pk} = \mathcal{E}_\mathrm{pk} \frac{c}{f_c} \sqrt{\frac{R_L}{Z_0}\frac{D}{4\pi}}
\end{equation}
where $c$ is the speed of light, $f_c=300$~MHz is the central frequency, weighted by the spectral shape of extensive air showers, $R_L=50 \ \Omega$ is the load impedance of the detector, $Z_0=377 \ \Omega$ is the impedance of free space, and $D=10$~dBi is the peak directivity of the ANITA horn antennas. The root-mean-square noise voltage is $\mathrm{V_{rms}}\sim 8.9 $~$\mu$V assuming a 290~Kelvin system noise temperature and 400~MHz bandwidth, corresponding to the band participating in the triggering. However, the smallest peak electric field value in the ANITA extensive air shower event ensemble was 446~$\mu$V/m corresponding to a peak voltage of 145~$\mu$V (or 18$\sigma$ above thermal noise)~\cite{Hoover_thesis}. This is due to the the multiple band coincidence design of the triggering system, which was not optimal for the extensive air shower pulse spectrum (see~\cite{Schoorlemmer_2015}). With the objective of setting an upper bound, we set trigger the threshold at half this value requiring that $\mathrm{V_{pk}}>72$~$\mu$V.


\section{Results}
\label{sec:upper_bound_results}
\begin{figure}[t!]
  \centering
   \includegraphics[page=1,trim = {0.8cm 1.5cm 0.9cm 1.8cm}, clip, width=0.32\linewidth]{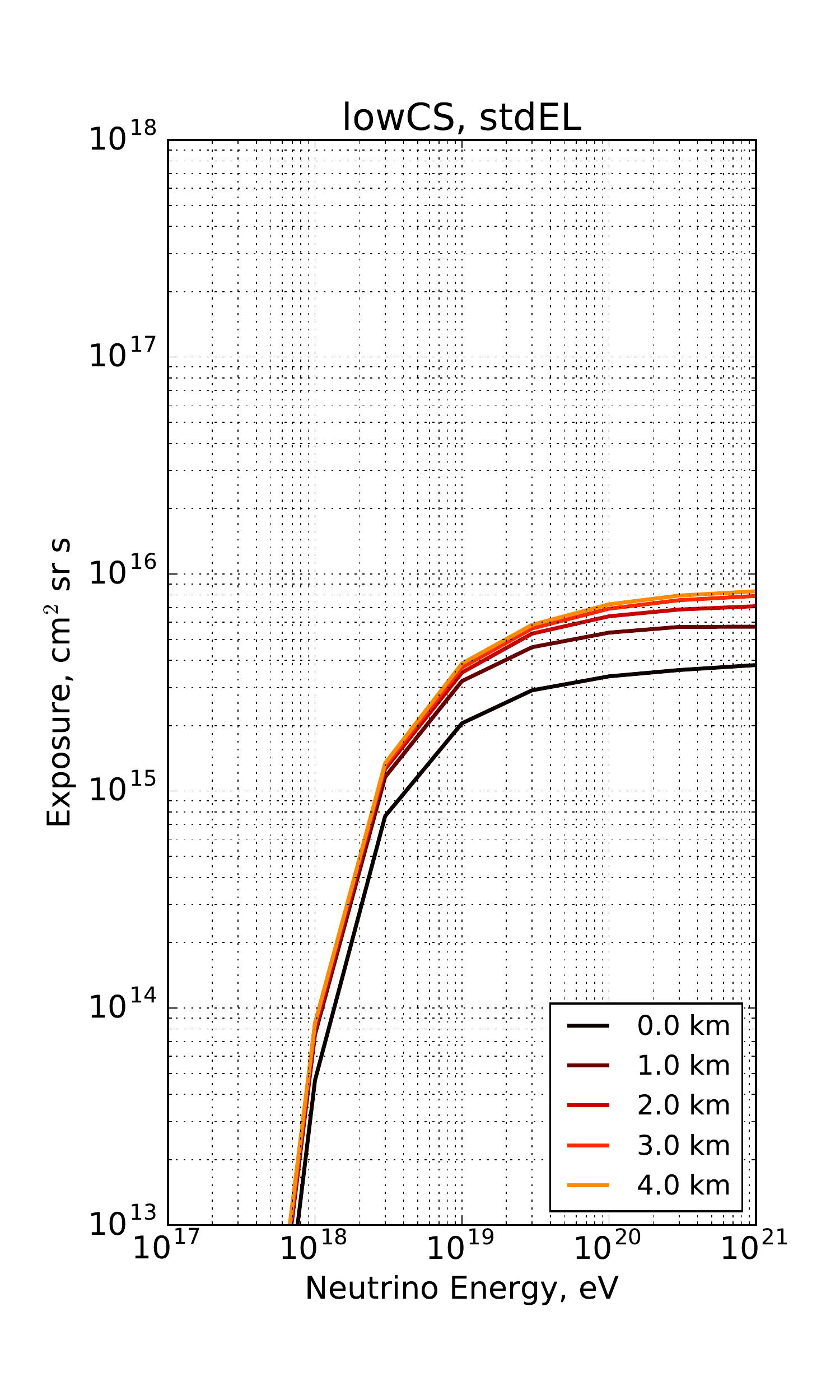} 
   \includegraphics[page=3,trim = {0.8cm 1.5cm 0.9cm 1.8cm}, clip,width=0.32\linewidth]{ana_sim_plots_143muV_thresh.pdf} 
   \includegraphics[page=5,trim = {0.8cm 1.5cm 0.9cm 1.8cm}, clip,width=0.32\linewidth]{ana_sim_plots_143muV_thresh.pdf} 
   \includegraphics[page=2,trim = {0.8cm 1.7cm 0.9cm 1.8cm}, clip,width=0.32\linewidth]{ana_sim_plots_143muV_thresh.pdf} 
   \includegraphics[page=4,trim = {0.8cm 1.7cm 0.9cm 1.8cm}, clip,width=0.32\linewidth]{ana_sim_plots_143muV_thresh.pdf} 
   \includegraphics[page=6,trim = {0.8cm 1.7cm 0.9cm 1.8cm}, clip,width=0.32\linewidth]{ana_sim_plots_143muV_thresh.pdf} 
   \caption{Upper bound on acceptance curves for ANITA assuming various ice shell thicknesses and models of the cross-section and energy loss. The ANITA upper bounds are compared to the Auger 2015~\cite{Auger_2015} and IceCube 2016~\cite{IceCube_2016} neutrino acceptances. Note that these are only a fair comparison for the standard neutrino cross-section and energy loss models (midCS, stdEL), otherwise the Auger and IceCube acceptance curves would also have to be modified.}
   \label{fig:acceptance}
\end{figure}

In Figure~\ref{fig:acceptance} we show upper bounds on the exposure of $\tau$-lepton-decay air showers of $\nu_\tau$ origin for the first flight of ANITA. The different panels, from left to right, correspond to the lower, middle, and upper ranges of the neutrino interaction cross-section within standard model uncertainties~\cite{Connolly_2011} (labeled lowCS, midCS, and uppCS, respectively). The top row of panels assume the standard $\tau$-lepton energy loss rates from the ALLM model~\cite{ALLM_97} and the bottom row assume the lower energy loss rates from the ASW model~\cite{ASW_2005} (labeled stdEL and lowEL, respectively). 

In each panel, we show the exposure corresponding to different ice sheet thicknesses. The presence of a layer of ice (or water) has a significant effect on the exposure, resulting in enhancements ranging between factors of $2-10$ depending on the energy and interaction models assumed within SM uncertainties.

In the top-middle panel of Figure~\ref{fig:acceptance}, corresponding to the middle cross-section curve and standard $\tau$-lepton energy loss rates, we show the ultra-high energy neutrino exposures of Auger~\cite{Auger_2015} and IceCube~\cite{IceCube_2016}. 
The upper bounds of the ANITA exposure curves are a factor of $\gtrsim60$ smaller than IceCube and Auger.  Given that neither IceCube or Auger have had a positive detection for neutrinos at ultra-high energies makes the diffuse-flux $\nu_\tau$ origin of the ANITA candidate event highly unlikely assuming these interaction models. However, it may have been due to a transient.

For panels in Figure~\ref{fig:acceptance} outside the standard values of the SM, the Auger and IceCube curves are not shown since, to our knowledge, neither experiment has published the dependence of their exposure on SM uncertainties of the neutrino interaction cross-section. The ANITA-1 exposure upper bounds do have a significant dependence on SM uncertainties, although not enough to make up for the factor of $\gtrsim60$ found in the standard values of the interaction models. However, without estimates of the Auger and IceCube exposures using the same interaction models, this is not a fair comparison.

\section{Discussion and Conclusion}
We have estimated an upper bound on the exposure of ANITA-1 to $\tau$-lepton decay air showers of $\nu_\tau$ origin. For standard values of the neutrino interaction cross-section and $\tau$ lepton energy loss rate, the exposure is $\gtrsim$60 times smaller than Auger and IceCube. Under these assumptions, the diffuse-flux $\nu_\tau$ origin of the $\tau$-lepton decay candidate event of ANITA-1 is highly unlikely. Although it may have potentially been due to a transient.

The ANITA $\nu_\tau$ exposure upper bound can change significantly depending on variations of the neutrino interaction cross-section and $\tau$-lepton energy loss rate within standard model uncertainties. However, it does not change significantly enough to make up for the tension with IceCube and Auger. Despite this, the comparison is not conclusive since neither IceCube or Auger have provided the exposure dependence on standard model uncertainties. 

Going beyond the standard model can result in models that significantly suppress or enhance the neutrino interaction cross-section~\cite{Cornet_2001, Jain_2002, Reynoso_2013}. The possibility that such models could result in the ANITA-1 $\tau$-lepton decay event candidate, without being in tension with IceCube and Auger, remains to be explored.

The origin of the ANITA-1 $\tau$-lepton event candidate remains a mystery. On-going analysis of the third and fourth flights of ANITA have the potential to confirm or falsify whether this event is of astrophysical origin.

\clearpage
\noindent{\it Part of this research was carried out at the Jet Propulsion Laboratory, California Institute of Technology, under a contract with the National Aeronautics and Space Administration.}

\end{document}